\documentstyle[emulateapj,epsfig]{article}
\newcommand{\ecm}{ergs$\cdot$s$^{-1}$cm$^{-2}$\AA$^{-1}$}
\newcommand{\msol}{$M_{\odot}$ }
\newcommand{\eps}{ergs$\cdot$s$^{-1}$}

\begin{document}
\title{The Compact UV Nucleus of M33}

\author{Guillaume Dubus}
\affil{Astronomical Institute, Univ. of Amsterdam, 
Kruislaan 403, 1098 SJ Amsterdam, the Netherlands}
\author{Knox S. Long}
\affil{Space Telescope Science Institute, 3700 San Martin Dr., Baltimore 
MD 21218}
\author{Philip A. Charles}
\affil{University of Oxford, Dept. of Astrophysics, Keble Road, Oxford 
OX1~3RH, UK}

\begin{abstract}
The most luminous X-ray source in the Local Group is associated with 
the nucleus of M33. This source, M33 X-8, appears modulated by $\sim$ 
20\% over a $\sim$ 106 day period, making it unlikely that the combined 
emission from unresolved sources could explain the otherwise persistent 
$\sim$10$^{39}$erg~s$^{-1}$ X-ray flux (\cite{dubus,hernquist}). We 
present here high resolution UV imaging of the nucleus with the 
{\it Planetary Camera} of the {\it Hubble Space Telescope} undertaken 
in order to search for the counterpart to X-8. The nucleus is bluer and
more compact than at longer wavelength images but it is still extended 
with half of its 3$\cdot$$10^{38}$ erg~s$^{-1}$ UV luminosity coming 
from the inner 0\farcs14. We cannot distinguish between 
a concentrated blue population and emission from a single object.
\end{abstract}


\section{Introduction}
The nearby galaxy M33 hosts the most luminous steady 
X-ray source in the Local Group, 
X-8. This source, with $L_X\sim$10$^{39}$erg~s$^{-1}$ (\cite{long}),
is coincident to within 5'' with the nucleus of the galaxy (\cite{schulman}).
Different models were invoked for X-8, including a quiescent mini-AGN
(\cite{trinchieri}; \cite{peres}), a collection of X-ray binaries 
(\cite{hernquist}) and a new type of X-ray binary (\cite{gott}).
Our ROSAT studies (\cite{dubus}) have shown that X-8
is very steady on both short and long time scales, except for low 
amplitude 
($\sim$20\%) variations which appear modulated on a $\sim$106 day period. 
This strongly favors a single source explanation for X-8.

We have interpreted the modulation in X-8 as ``superorbital'', similar to
that seen in a number of bright galactic X-ray binaries which were monitored
by e.g. the Vela 5B satellite (\cite{smale}).  X-8 is then likely to be a 
$\geq$10$M_\odot$ black-hole X-ray binary
(the high mass is required to account for the observed luminosity) but with
a companion in a much shorter orbital period than 106d. This is supported by
the extremely low velocity dispersion of the nucleus which limits the mass 
of a 
central black hole in M33 to $\leq$5$\times$10$^4$M$_\odot$ (\cite{kormendy}, KM93).
This and the high central stellar density imply that the nucleus
is an extremely relaxed, post core-collapse stellar system 
(comparable for instance to a galactic globular cluster such as M15).
 A significant number of stellar collisions/interactions could have taken 
place, eventually leading to the creation of exotic
interacting binaries (e.g. \cite{hut}).

The next step toward unravelling the mystery of X-8 would be to
identify its optical counterpart.  However, even with  optimistic 
$L_X/L_{opt}$ ratios for either X-ray binaries or AGN the counterpart 
would only have V$\sim$21 compared to a core brightness of V$\sim$14.
But with the optical spectral type of an F supergiant, the dominance of 
the M33 visual core cannot extend to UV wavelengths
where the hot/flat spectrum of X-8's associated disc ought to be a
significant contributor.  This is true despite 
evidence for a color gradient in the nucleus (KM93, \cite{mighell,lauer}) 
suggesting that a period of recent star formation has taken place and/or
that collisions have modified the central star population.  Here we report
an attempt to find the counterpart using the UV imaging capabilities of HST.

\section{Observations and Reduction}
Observations were carried out with the HST WFPC-2 on June 12, 1997 using 
three different filters. The nucleus was positioned at the center of the 
{\it Planetary Camera} ($\alpha$ 1:33:51.1, $\delta$ 30:39:39, J2000). 
During the first orbit two 1200s exposures were made with the F160BW 
(`UV' filter, $\bar{\lambda} 1491$). In the following orbit, two 800s 
exposures were made with the F300W filter (`U' filter, $\bar{\lambda} 2942$\AA)
and one 500s exposure with the F439W filter (`B' filter, $\bar{\lambda} 4300$\AA). All these exposures were made with the gain setting at 7.

In addition, we have extracted recalibrated archival data from the Space 
Telescope European Coordinating Facility (ST-ECF) Archive. These data 
included two 40s exposures with the F555W at gain 14 (`V' filter, $\bar{\lambda} 5397$\AA), two 40s exposures with the F814W at gain 14 (`I' filter, $\bar{\lambda} 7924$\AA) and six 300s exposures with the F1042M at gain 7 ($\bar{\lambda} 10190$\AA) filters, all centered on the nucleus and dating from September 26-27, 1994. The archival V and I data were previously discussed by \cite{lauer} (L98). Fig. 1 shows the central region of the reduced UV, U and B images.

\begin{figure}
\plotone{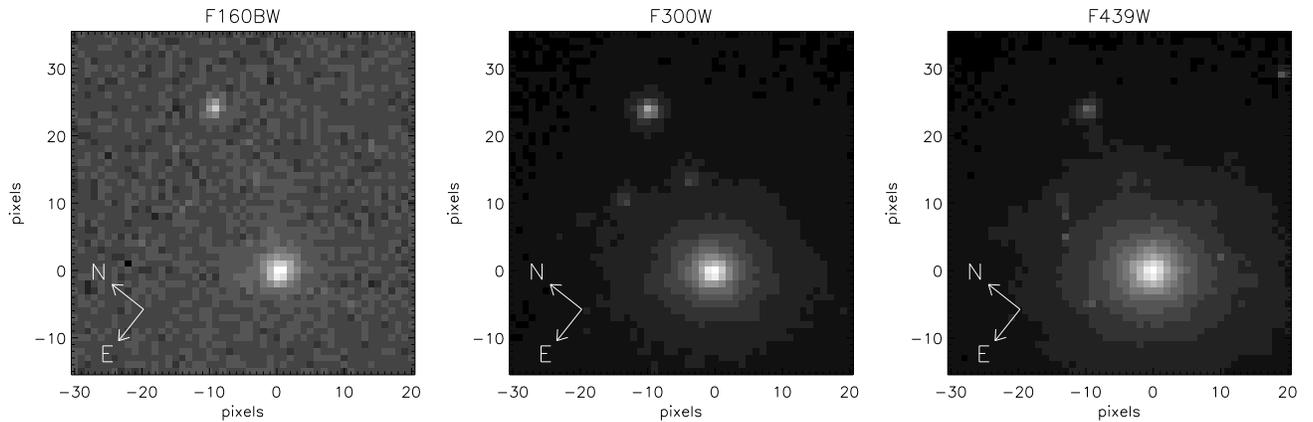}
\caption{Central portion of the Planetary Camera showing the nucleus 
and the nearby star in the F160BW `UV', F300W `U' and F439W `B' filters. Each pixel is 0.0455'' i.e. each image is about 2.3'' $\times$ 2.3''.} 
\label{figure1}
\end{figure}

The data were reduced using the HST calibration pipeline. The signal-to-noise ratio of the only existing F160BW flat was quite low. Following advice from the WFPC-2 group at STScI, we decided to use the F255W flat field. Effects of cosmic rays were reduced on those images for which we had multiple exposures with the {\it IRAF stsdas} routine {\it crrej}. Images of the nuclear region as observed through the F160BW, F300W and F439W filter with the PC are shown in Fig. 1.
\placefigure{figure1}

\section{Analysis}
\begin{figure}
\plotone{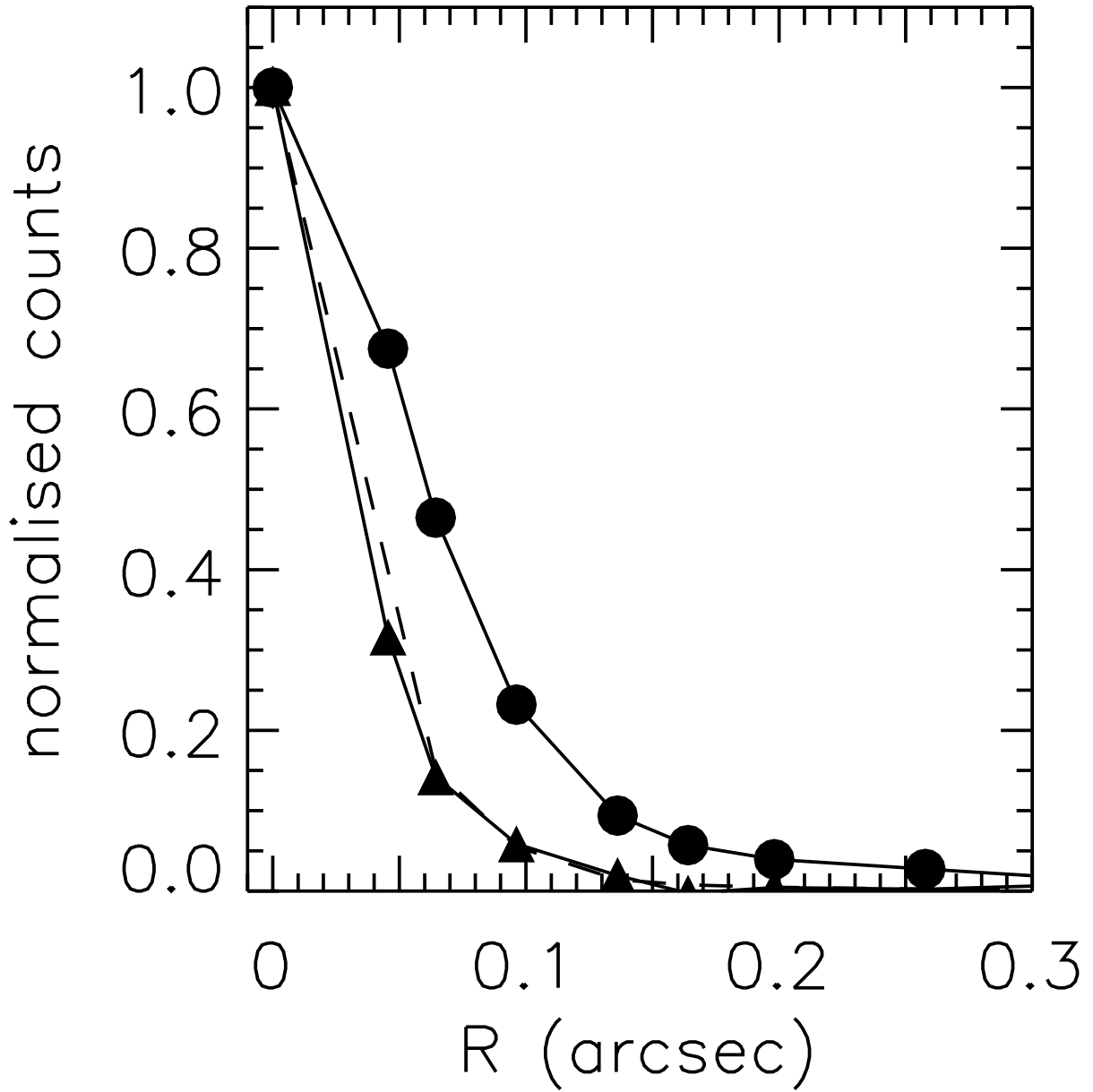}
\caption{Radial profiles of the nucleus and nearby star in UV. The nucleus is shown by the filled circles + continuous line. The star NNW of the nucleus (see Fig. 1) is shown by the triangles + continuous line. The best fitting PSF to the star is shown by the dashed line. The nucleus in UV is clearly extended.}
\label{figure2}
\end{figure}

Our values for the total flux of the nucleus in the different bands agree with those of \cite{gordon}. The radial profiles of the B, V and I data are also consistent with KM93 and L98. 
As is apparent from Fig. 1, the nucleus of M33 appears more concentrated
in the F160BW filter than in the longer wavelength filters.  Nevertheless,
as shown in Fig. 2, the profile in the F160BW image of the nucleus is
extended compared to the profile of the star located about 1" NNW 
from the nucleus
and to PSF profiles calculated using the HST PSF generating routine TinyTim 
(\cite{krist}). Further investigation of the radial structure of the UV emission calls for deconvolution of the data taking into account the different instrumental effects in the {\it Planetary Camera} (L98). Since the UV image does not have a large enough signal-to-noise to allow for a proper deconvolution, we chose instead to fit convolved models to the data.
\placefigure{figure2}

\subsection{Fits of extended emission models}
Following KM93, we fitted radial profile models of the form:
\begin{equation}
\Sigma=\Sigma_{\rm o} \left(1+(r/r_{\rm o})^2\right)^{-n}
\end{equation}
The model and the PSF are oversampled on a 4x4 grid for each PC pixel. For a given set of $(r_{\rm o},n)$, the convolved model is moved on the 4x4 grid, rebinned to the PC resolution and compared to the data. The comparison with the data is performed in a 64x64 pixel aperture (about 3''x3'') but we have verified that larger and smaller apertures gave similar results. We have assumed an A type spectrum for the PSF but other choices do not affect our conclusions. The parameter $r_{\rm o}$ is varied between 0.05 and 2 PC pixels and $n$ is varied between 0.5 and 2. The results from the $\chi^2$ minimization routine are presented in Tab. 1 and Fig. 3. The quoted errors correspond to 10\% higher values than at the minimum of the fit function. In the noisy UV band only lower bounds to the parameters could be extracted. Here we give errors on the FWHM instead of on $r_{\rm o}$.
\placetable{table1}

\begin{figure}
\plotone{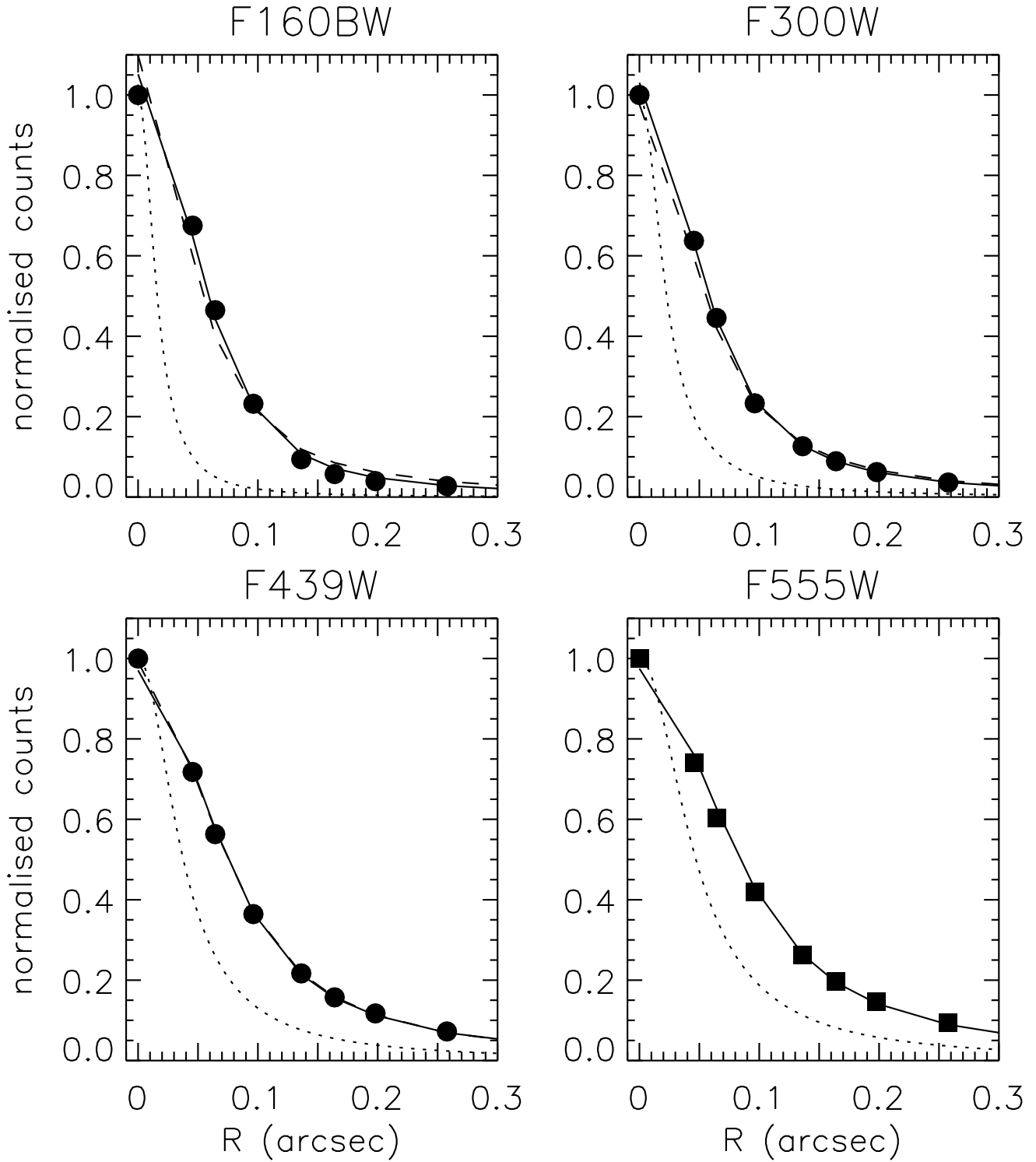}
\caption{Radial profiles of the nucleus in the different filters (data points). The best fitting PSF convolved models assuming only extended emission (paragraph 3.1) are shown by straight lines. The unconvolved models (Eq. 1) are shown as dotted lines. PSF convolved models assuming extended emission and a point source (paragraph 3.2) are shown with dashed lines in the F160BW (UV), F300W (U) and F439W (B) profiles. Normalisation factors are (in Mag/arcsec$^{2}$) 11.67 (F160BW), 11.65 (F300W), 12.10 (F439W) and 11.76 (F555W).}
\label{figure3}
\end{figure}

The FWHM of the models decreases in accordance with the blue colour gradient. In all cases $n\approx$ 1 suggesting that the distribution of light at large radii is the same in all the filters. This is consistent with the flat colour profiles observed for $r\gg r_{\rm o}$. With $n$ fixed at 0.75 as in L98 we also find a FWHM for the F555W data of 0\farcs07. This is clearly not the best solution when $n$ varies (Tab. 1). We find a higher FWHM (0\farcs09). We note that KM93 find $n$ between 0.8--1.3 and a FWHM below 0\farcs1. The similar values found for $r_{\rm o}$ in the V, I, F1042M and (to a lesser extent) B bands indicate that their radial profiles are comparable (i.e. the colour gradients are much reduced between those bands than when compared to the UV and U so that, for example, on first approximation the {\it V-I} colour gradient is negligible when compared to {\it UV-V}).
\placefigure{figure3}

\subsection{Fits with an additional point source}
Since the V, I and F1042M show very close light distributions, we investigated whether the compact emission from the UV and U filter could be explained by an underlying extended population having the V band distribution plus a blue point source. We fixed $r_{\rm o}$=1 PC pixel, $n$=1 and superposed at the center of this model a point source of varying relative strength. For the V and redward filters, the best fits were consistently obtained with a nil contribution from the point source. However, a point source of increasing strength was needed in B, U and UV. Only those models with fit values within the errors of the previous extended emission fits (Tab. 1) were kept i.e. the fits here are as good or better than the previous ones (see Fig. 3). The contribution of the point source to the total flux within a 1\farcs45 circular aperture is summarized in Tab. 2. The corresponding magnitudes are given in the VegaMAG system using updated tables for the zeropoints (\cite{holtzman}).
\placetable{table2}

The best fits for the B, U and UV are shown in Fig. 3 by dashed lines. They are indistinguishable from the extended emission fits. As a result we cannot, based on the data in hand, distinguish between a model in which emission is extended in the UV but has a smaller core radius than at longer wavelengths and a composite model consisting of a point source and an underlying distribution characterised by the visible light profile.

\section{Discussion}
The nucleus of M33 has a composite spectrum ranging from A7V at $\lambda$$\sim$3800\AA{} to F5V at $\lambda$$\sim$4300\AA{} (\cite{oconnell}). This requires at least a two component population in most models with the blue emission being due to young metal-rich stars. The colour gradient in {\it B-R} implies this young population is more centrally condensed. The nucleus could have been the site of episodic starbursts with the youngest stars being about 10 Myr old (\cite{oconnell,vand,schmidt} and references therein). Recently \cite{gordon} have argued that a single 70 Myr old starburst reproduces the UV to IR spectral energy distribution within 4\farcs5 of the center if dust is correctly taken into account\footnote{\cite{gordon} propose that X-8 is a high mass X-ray binary with an early B companion. But as had been noted by \cite{oconnell}, the high mass tranfer rate needed to power the 10$^{39}$ \eps{} luminosity implies an uncomfortably short evolutionary time-scale ($\sim 10^5$ years).}. The nucleus is also very similar to a globular cluster and is likely to have undergone core-collapse (\cite{hernquist}, KM93). Blue stars formed in collisions at the center might explain the colour gradient. This model may have difficulties accounting for the UV luminosity (\cite{hernquist}, L98, \cite{gordon}).

\cite{massey} detected the nucleus in the UV but with 5'' resolution. Hence they had proposed that the blue component of the nucleus could be due to unresolved emission from a few hot stars. However, the HST data shows the UV emission is very compact with no stars of comparable brightness within 5'' of the center. The total flux is about 6.3$\cdot$10$^{-15}$ \ecm{} in a 4\farcs55 aperture or about 2.8$\cdot$10$^{38}$ \eps at 800 kpc. From the best fit model, $\sim$50\% of the UV light comes from the inner 0\farcs14 of the nucleus. Any model of the structure of the nucleus has to explain this UV emission from a region only $\sim$ 0.55 pc across. If a point source is present, this source is responsible for $\sim$30\% of the UV flux within 0\farcs14. The contribution from the underlying extended population is subsequently reduced.

The nucleus of NGC 205, at a comparable distance of 720kpc, is in many ways similar to that of M33, but without the X-ray source. It is globular cluster like, has a comparable $M_V$ and a low upper limit of 9$\cdot$10$^{4}$\msol on the mass within the central pc (\cite{heath}). These authors find that the nucleus is more extended, with a F555W FWHM of 0\farcs2, $F_{\rm F555W}$=1.8$\cdot$10$^{-15}$ \ecm{} and only $F_{\rm F160BW}$=6.0$\cdot$10$^{-16}$ \ecm{} (within 0\farcs273). Using the same aperture on the M33 data we find for M33 $F_{\rm F555W}$=3.4$\cdot$10$^{-15}$ \ecm{} and $F_{\rm F160BW}$=2.5$\cdot$10$^{-15}$ \ecm{}. Excluding the contribution from the point source, the F160BW flux of M33 is about 1.5$\cdot$10$^{-15}$ \ecm{} and the ratios of the fluxes between the bands become comparable.

If there is in fact a point source at the center of the nucleus of M33, the magnitude estimates in Table 2 are consistent with a Rayleigh-Jeans tail. We estimate the total UV flux of the source would be $\sim$1.0$\cdot$10$^{-15}$ \ecm{}, which would suggest $L_{\rm opt}$/$L_X$$\sim$0.05 which is reasonable for the type of X-ray source we have postulated X-8 to be. This source would be responsible for most of the colour gradient in {\it UV-B} and {\it U-B} and thus the very compact appearance of the nucleus at these wavelengths. As it contributes only $\sim$18\% of the total F160BW flux and $\sim$8\% of the F300W flux, the spectral energy distribution of the nucleus is not changed much and the consequences on population synthesis studies should be minor. However, it does have a major influence in that it relaxes the constraints on e.g. mass segregation to explain the very strong colour gradients in {\it UV-B} and {\it U-B}. 

The star located about 1" from the nucleus in M33 has $B$=19.45, $V$=19.25, $M_{\rm F300W}$=18.05 and $M_{\rm F160BW}$=17.60. The count rates are compatible with an A0 type spectrum which would make it similar, although fainter, to the two A supergiants detected by \cite{massey}. The fluxes in the F300W and F160BW bands, at the distance of M33 ($\sim$800 kpc), are $\sim$ $10^{37}$ erg$\cdot$s$^{-1}$. The positional accuracy of the {\it ROSAT HRI} does not rule out this star as a possible counterpart to the X-ray source X-8. The ratio $L_X$/$L_{\rm opt} \sim$ 100 and the observed $U-B$ and $B-V$ would agree with what is expected from a low-mass X-ray binary (\cite{vanp}). But its absolute magnitude ($M_V\approx$-5.2) would make it more similar to a high mass X-ray binary. The main argument against this star being the X-ray binary is X-8's unique character and special location at the nucleus.  Given that \cite{massey} found $\sim$300 analogous UV sources in M33, it would be remarkable that the one near the nucleus is the most luminous X-ray source in the Local Group.

\section{Conclusion}
The UV high resolution images obtained with the HST {\it Planetary Camera} show the nucleus of M33 is extremely compact. We have fitted convolved models to the radial profiles in the different bands from which we find the FWHM of the nucleus in UV to be $\sim$0\farcs035 and $\sim$0\farcs090 in V. About half of the UV flux comes from the inner 0\farcs14 of the nucleus. The UV and U profiles are also well fitted if one assumes a blue point source superposed on an extended population with the same FWHM as in V. If this is the correct model for the nucleus, then this point source is likely to be the UV counterpart to the very luminous X-ray source X-8. Such a counterpart would be responsible for most of the strong colour gradient seen in UV. Its contribution to the total UV flux of the nucleus would be about 18\%. Models for the structure of the nucleus still need to account for the remainder of the UV flux but the constraints on population segregation (more compact blue star population) are reduced. High spatial resolution UV spectroscopy of the nucleus is the obvious next step, which we will undertake shortly.

\acknowledgments
We thank Sylvia Baggett for help with the F160BW filter reduction.
G.D. and K.S.L wish to thank the Oxford Astrophysics Department where
part of this work was completed.
We acknowledge support by the British-French joint research
program {\it Alliance} and by NASA grant NAG 5-1539 to the STScI.
Based on observations with the NASA/ESA Hubble Space Telescope obtained at 
the STScI which is operated by the AURA under NASA contract No. NAS5-26555.

\begin{table}
\begin{tabular}[h]{lcll}
\multicolumn{4}{c}{Table 1: Fits of extended emission models}\\
\hline
\hline
Filter & $r_o$ & $n$ & FWHM\\
& {\small (PC pix.)} & & {\small (0\farcs001)} \\
\hline
F160BW & 0.4 & 1.1$_{-0.2}$          & 35$_{-25}$\\
F300W  & 0.5 & 1.0$^{+0.2}_{-0.1}$   & 45$_{-15}^{+25}$\\
F439W  & 0.8 & 1.0$^{+0.1}_{-0.1}$   & 75$_{-15}^{+20}$\\
F555W  & 1.0 & 1.0$^{+0.05}_{-0.05}$ & 90$_{-5}^{+15}$\\
F814W  & 0.9 & 0.9$^{+0.1}_{-0.05}$  & 90$_{-15}^{+15}$\\
F1042M & 1.0 & 0.9$^{+0.1}_{-0.1}$   & 100$_{-25}^{+15}$\\
\hline
\label{table1}
\end{tabular}
\end{table}

\begin{table}
\begin{tabular}[h]{llc}
\multicolumn{3}{c}{Table 2: Fits with additional point source}\\
\hline
\hline
%\multicolumn{3}{l}{Ext. emission model : $r_o$=1 {\small PC pixel}, $n$=1}\\
%\hline
Filter & \multicolumn{2}{l}{Point source flux \& magnitude}\\
 &  \multicolumn{2}{l}{\small within 1\farcs45 aperture}\\
& {\small (\%)} & {\small (VEGAmag)} \\
\hline
F160BW &  18$^{+4}_{-3}$ & 16.9 \\
F300W  &  8$^{+4}_{-2}$ & 17.5 \\
F439W  &  1$^{+2}_{-1}$ & 19.7 \\
\hline
\end{tabular}
\label{table2}
\end{table}

\begin{thebibliography}{}

\bibitem[van den Bergh 1991]{vand} 
van den Bergh S., 1991, \pasp, 103, 609

\bibitem[Biretta et al. 1996]{wfpc2}
Biretta J.A., et al., 1996, The WFPC-2 Instrument Handbook, version 4.0, STScI

\bibitem[Dubus et al. 1997]{dubus} 
Dubus G., Charles P.A., Long K.S., \& Hakala P.J., 1997, \apjl, 490, 47

\bibitem[Gordon et al. 1999]{gordon}
Gordon K.D., Hanson M.M., Clayton G.C., Rieke G.H., \& Misselt K.A., 1999, \apj, in press

\bibitem[Gottwald, et al. 1987]{gott} 
Gottwald M., Pietsch W., \& Hasinger G., 1987, A\&A, 175, 45

\bibitem[Heath Jones et al. 1996]{heath}
Heath Jones D., \& al., 1996, \apj, 466, 742

\bibitem[Hernquist et al. 1991]{hernquist} 
Hernquist L., Hut P., \& Kormendy J., 1991, Nature, 354, 376

\bibitem[Holtzman et al. 1995]{holtzman} 
Holtzman J.A., Burrows C.J., Casertano S., Hester J.J., Trauger J.T., Watson A.M., \& Worthey G., 1995, \pasp, 107, 1065

\bibitem[Hut et al. 1992]{hut} 
Hut P., McMillan S., Goodman J., Mateo M., Phinney E.S., Pryor C., Richer H.B., Verbunt F., \& Weinberg M., 1992, \pasp, 104, 981

\bibitem[Kormendy \& McClure 1993]{kormendy} 
Kormendy J., \& McClure R.D., 1993, \aj, 105, 1793  (KM93)

\bibitem[Krist \& Hook 1997]{krist} 
Krist J., \& Hook R., 1997, The Tiny Tim User's Guide, version 4.4

\bibitem[Lauer et al. 1998]{lauer} 
Lauer T.R., Faber S.M., Ajhar E.A., Grillmair C.J., \& Scowen P.A., 1998, \aj, 116, 2263 (L98)

\bibitem[Long et al. 1996]{long} 
Long K.S., Charles P.A., Blair W.P., \& Gordon S.M, 1996, \apj, 466, 750

\bibitem[Massey et al. 1996]{massey} 
Massey P., Bianchi L., Hutchings J.B., \& Stecher T.P., 1996, \apj, 469, 629

\bibitem[Mighell \& Rich 1995]{mighell} 
Mighell K.J., \& Rich R.M., 1995, \aj, 110, 1649

\bibitem[O'Connell 1983]{oconnell}
O'Connell R., 1983, \apj, 267, 80

\bibitem[van Paradijs \& McClintock 1995]{vanp}
van Paradijs J., \& McClintock J.E., 1995, chapter 2 in {\it X-ray binaries}, ed. W.H.G Lewin, J. van Paradijs \& E.P.J. van den Heuvel, Cambridge University Press

\bibitem[Peres et al. 1989]{peres}
Peres G., Reale F., Collura A., \& Fabbiano G., 1989, \apj, 336, 140

\bibitem[Schmidt et al. 1990]{schmidt}
Schmidt A.A., Bica E., \& Alloin D., 1990, \mnras, 243, 620

\bibitem[Schulman \& Bregman 1995]{schulman}
Schulman E., \& Bregman J.N., 1995, ApJ, 441, 568 

\bibitem[Smale \& Lochner 1992]{smale} 
Smale A.P., \& Lochner J.C., 1992, \apj, 395, 582

\bibitem[Trinchieri et al. 1988]{trinchieri} Trinchieri G., Fabbiano G., \& Peres G.,1988, \apj, 325, 531

\end{thebibliography}
\end{document}